\documentstyle[prl,aps,psfig,floats,preprint]{revtex}

\begin{document}
\def\beq{\begin{equation}}
\def\eeq{\end{equation}}
\def\ber{\begin{eqnarray}}
\def\eer{\end{eqnarray}}
\def\apj{{Astroph.\@ J.\ }}
\def\mn{{Mon.\@ Not.@ Roy.\@ Ast.\@ Soc.\ }}
\def\asta{{Astron.\@ Astrophys.\ }}
\def\aj{{Astron.\@ J.\ }}
\def\prl{{Phys.\@ Rev.\@ Lett.\ }}
\def\prd{{Phys.\@ Rev.\@ D\ }}
\def\nucp{{Nucl.\@ Phys.\ }}
\def\nat{{Nature\ }}
\def\plb {{Phys.\@ Lett.\@ B\ }}
\def \jetpl {JETP Lett.\ }
\def\etal{{\it et al.}}
\def\ie {{\it ie~}}
\def\half{{1\over 2}}

\def\lsim{\
  \lower-1.5pt\vbox{\hbox{\rlap{$<$}\lower5.3pt\vbox{\hbox{$\sim$}}}}\ }
\def\gsim{\
  \lower-1.5pt\vbox{\hbox{\rlap{$>$}\lower5.3pt\vbox{\hbox{$\sim$}}}}\ }

\title{
New Vistas in Braneworld Cosmology\footnotetext{\em Honorable mention in the
2002 Essay Competition of the Gravity Research Foundation.}}

\author{Varun Sahni$^a$ and Yuri Shtanov$^b$}
\address{$^a$Inter-University Centre for Astronomy and Astrophysics, \\ Post Bag 4,
Ganeshkhind, Pune 411~007, India \\
$^b$Bogolyubov Institute for Theoretical Physics, Kiev 03143, Ukraine}

\maketitle

\thispagestyle{empty}

\begin{abstract}
Traditionally, higher-dimensional cosmological models have sought to provide a
description of the fundamental forces in terms of a unifying geometrical
construction. In this essay we discuss how, in their present incarnation,
higher-dimensional `braneworld' models might provide answers to a number of
cosmological puzzles including the issue of dark energy and the nature of the
big-bang singularity.
\end{abstract}

\newpage

\section*{Compact extra dimensions}

The idea that physical space has more than three dimensions is not new.
Nordstr\"om (1914), and independently Kaluza (1921) and Klein (1926) suggested
the presence of a fifth spatial dimension in their attempts to unify gravity
with the electromagnetic force \cite{kk}. The flowering of Kaluza-Klein
cosmology took place in the 1980's and was associated with the development of
non-Abelian gauge field theories. The major aim of this program was to recover
the gauge symmetries of the standard model from compact (hidden) dimensions,
and it was assumed that the compactification scale ${\cal R}$ was of the same
order as the Planck length: ${\cal R} \sim \ell_P \simeq 10^{-33}$~cm. A fresh
impetus to this paradigm was given in \cite{dvali}, where it was suggested that
compact extra dimensions may be macroscopic, ${\cal R} \lsim 1$~mm, while our
space-time is described as a lower-dimensional domain wall (brane) where all
the matter is concentrated. Within this setting, the fundamental
higher-dimensional Planck scale could become as small as $\sim$~1~TeV, thus
ridding particle physics of the hierarchy problem. (The hierarchy problem
arises because the Planck scale is so much higher than the known scales in
physics, for instance $M_P/M_W \sim 10^{17}$, where $M_W$ is the mass of the
vector boson which mediates the electroweak force.) The reasoning behind this
scenario becomes clear when one considers two test masses $m_1$\,, $m_2$
separated by a distance $r \ll {\cal R}$ in a $(4+n)$-dimensional universe and
interacting via the Newtonian potential
\begin{equation} \label{pot1}
V(r) \sim \frac{m_1\, m_2}{M^{n+2}_N}\, \frac{1}{r^{n+1}} \quad (r \ll {\cal
R}),
\end{equation}
where $M_N$ is the $N$-dimensional Planck mass. If the same two particles are
placed much further apart, then, because the gravitational field lines
associated with $m_1$\,, $m_2$ cannot penetrate into the extra dimensions at
such large distances, the potential at large separations becomes
\begin{equation} \label{pot2}
V(r) \sim \frac{m_1\, m_2}{M^{n+2}_N {\cal R}^n}\, \frac{1}{r} \sim \frac{m_1\,
m_2}{M_4^2}\, \frac{1}{r} \quad (r \gg {\cal R}).
\end{equation}
From (\ref{pot1}) \& (\ref{pot2}) one finds that the effective four-dimensional
Planck mass is simply given by $M^{2}_4 \sim M^{n+2}_N {\cal R}^n$.
Substituting ${\cal R} \sim 1$~mm and $n = 2$, one finds $M_N \sim 1$~TeV.

\section*{Noncompact extra dimension}

A novel approach to higher-dimensional cosmology emerged when Randall and
Sundrum postulated the existence of a {\em noncompact\/} spacelike fifth
dimension \cite{rs99}. According to this world-view, our perception of `normal'
four-dimensional gravity arises because we live on a domain wall (brane)
embedded in or bounding a `bulk' anti-de~Sitter space (AdS)\@. The metric
describing the full (4+1)-dimensional space-time is non-factorizable, and the
small value of the true five-dimensional Planck mass is related to its large
effective four-dimensional value by the extremely large warp of the
five-dimensional space. This new approach to extra dimensions has far-reaching
cosmological implications, some of which will be examined in this essay.

A general framework for braneworld cosmology in which the brane is the boundary
of the five-dimensional bulk is provided by the action
\begin{equation} \label{action}
S = M^3 \left[\int_{\rm bulk} \left( R_5 - 2 \Lambda_{\rm b} \right) - 2
\int_{\rm brane} K \right] + \int_{\rm brane} \left( m^2 R_4 - 2 \sigma \right)
+ \int_{\rm brane} L \left( h_{\alpha\beta}, \phi \right) \, .
\end{equation}
Here, $R_5$ is the scalar curvature of the metric $g_{ab}$ in the
five-dimensional bulk, and $R_4$ is the scalar curvature of the induced metric
$h_{\alpha\beta}$ on the brane. The quantity $K = K_{\alpha\beta}
h^{\alpha\beta}$ is the trace of the extrinsic curvature $K_{\alpha\beta}$ on
the brane defined with respect to its {\em inner\/} normal. $L
(h_{\alpha\beta}, \phi)$ is the four-dimensional matter field Lagrangian, $M$
and $m$ denote, respectively, the five-dimensional and four-dimensional Planck
masses, $\Lambda_{\rm b}$ is the bulk cosmological constant, and $\sigma$ is
the brane tension.  Integrations in (\ref{action}) are performed with respect
to the natural volume elements on the bulk and brane.

In virtually all the braneworld models, the extra dimension is assumed to be
spacelike, so that the brane is embedded in a Lorentzian five-dimensional
manifold.  However, no physical principle appears to prevent us from
considering a complementary model in which the extra dimension is timelike, so
that the brane is a Lorentzian boundary of a five-dimensional space with
signature $(-,-,+,+,+)$ (see \cite{Kofinas}). This model is described by action
(\ref{action}) in which $K_{\alpha\beta}$ is the extrinsic curvature defined
with respect to the {\em outer\/} normal to the brane. Both models will be
examined in this essay.

Action (\ref{action}) leads to the following general cosmological equation (see
\cite{Shtanov1}):
\begin{equation} \label{cosmo}
m^4 \left( H^2 + {\kappa \over a^2} - {\rho + \sigma \over 3 m^2} \right)^2 =
\epsilon M^6  \left(H^2 + {\kappa \over a^2} - {\Lambda_{\rm b} \over 6} - {C
\over a^4} \right) \, ,
\end{equation}
where $\epsilon = 1$ if the extra dimension is spacelike, and $\epsilon = -1$
if it is timelike. The constant $C$ is the integration constant that, in the
case $\epsilon = 1$, corresponds to the black-hole mass of the
Schwarzschild--AdS solution in the bulk, $H \equiv \dot a/a$ is the Hubble
parameter on the brane, and $\kappa = 0, \pm 1$ corresponds to the sign of the
spatial curvature on the brane.

The Randall--Sundrum model assumes $m = 0$, and the cosmological brane equation
(\ref{cosmo}) in this case reduces to
\begin{equation}
H^2 + \frac{\kappa}{a^2} = \frac{8\pi}{3 M_4^2} \left( \rho +
\frac{\rho^2}{2\sigma} \right) + \frac{\Lambda}{3} + \frac{C}{a^4} \, ,
\label{eq:brane}
\end{equation}
where
\begin{equation} \label{constants}
M_4^2 = \frac{12 \pi \epsilon M^6}{\sigma} \, , \qquad \Lambda =
\frac{\Lambda_{\rm b}}{2} + \frac{\epsilon \sigma^2}{3 M^6} \, .
\end{equation}
A cosmologically viable model requires $M_4^2 > 0$, hence, $\epsilon/\sigma >
0$. The term $C / a^4$ is called `dark radiation.'

The additional term in the parentheses of (\ref{eq:brane}),
$\rho^2/2\sigma$, can be of great relevance for inflationary
model building on the brane.
Consider, for example, a scalar field propagating on the brane
that satisfies
\begin{equation}
{\ddot \phi} + 3H {\dot \phi} + V'(\phi) = 0\, ,
\label{eq:kg}
\end{equation}
where $H$ is given by (\ref{eq:brane}) with $\rho = \half{\dot\phi}^2 +
V(\phi)$. For the usual case $\epsilon = 1$ ($\sigma > 0$), the new term
$\rho^2/2\sigma$ in (\ref{eq:brane}) {\em increases\/} the damping experienced
by the scalar field as it rolls down its potential. This is reflected in the
slow-roll parameters \cite{maartens} (derived below for the case when $\sigma /
V \ll 1$)
\begin{equation}
\zeta \simeq \frac{M_4^2}{4\pi} \left (\frac{V'}{V}\right )^2
\frac{\sigma}{V}\, , \qquad \eta \simeq \frac{M_4^2}{4\pi}\, \frac{V''}{V}\,
\frac{\sigma}{V}\, . \label{eq::slow-roll}
\end{equation}
At early times, one expects $\sigma / V \ll 1$, and the slow-roll condition
$\zeta, \eta \ll 1$ becomes surprisingly easy to achieve. As a result,
inflation can be driven by {\em steep potentials\/}, including some ($V \propto
\phi^{-\alpha}$ etc.\@) not normally associated with inflation \cite{copeland}.
Thus, the class of potentials which give rise to inflation vastly increases,
and the possibility of realising inflation becomes that much easier in
braneworld cosmology.

Braneworld inflation leaves behind an imprint on the cosmological gravity-wave
background by increasing its amplitude and creating a distinct `blue tilt' in
its spectrum. Braneworld models therefore allow the possibility of being tested
through future LISA-type searches for gravity waves \cite{sami}.

As demonstrated in \cite{brane1}, the presence of the $\rho^2$ term in the RHS
of the equations for the braneworld persists when one considers anisotropic
space-times such as Bianchi~I and results in a fundamental change in the nature
of the big-bang singularity, which is shown to be matter dominated and {\em
isotropic\/}. This result marks a big departure from our conventional picture
of a cosmological singularity which, within the general-relativistic setting,
is expected to be vacuum dominated and significantly anisotropic \cite{mac}.

The $\rho^2$ term in (\ref{eq:brane}) plays an even more interesting role if
the extra dimension is timelike. In this case, we have $\epsilon = -1$, and the
requirement $M_4^2 > 0$ implies a negative brane tension, $\sigma < 0$. As a
result, the $\rho^2$ term comes with negative sign in Eq.~(\ref{eq:brane}), and
one finds that instead of becoming singular, the universe bounces at high
densities when $\rho/2|\sigma| \sim 1$. Thus the initial `big-bang singularity'
may be avoided in a braneworld which has a timelike extra dimension~!

\section*{Braneworld dark energy}

More surprises appear if we enlarge the scope of braneworld cosmology and
permit $m\neq 0$ in action (\ref{action}). This means that we must now consider
the contribution from the term $\displaystyle m^2 \int_{\rm brane} R_4$ which
arises when one incorporates quantum effects generated by matter fields
residing on the brane. (The origin of this idea goes back to Sakharov
\cite{Sakharov}, who first suggested the possibility of `inducing gravity'
through the backreaction of matter fields in the action.) The resulting field
equations on the brane can be written as
\begin{equation} \label{brane}
m^2 G_{\alpha\beta} + \sigma h_{\alpha\beta} = T_{\alpha\beta} + \underline{M^3
\left(K_{\alpha\beta} - h_{\alpha\beta} K \right)} \, ,
\end{equation}
where $T_{\alpha\beta}$ is the stress-energy tensor on the brane, and the
underlined term makes braneworld dynamics different from general relativity.
We see that a new length scale $\ell = 2 m^2 /M^3$ emerges in the theory
\cite{DGP}. On dimensional grounds, $G_{\alpha\beta} \sim
\left(K_{\alpha\beta}\right)^2 \sim 1/r^2$, and an examination of (\ref{brane})
reveals that braneworld dynamics approaches general relativity (GR) {\em on
small scales\/} $r \ll \ell$. Departure from GR in this braneworld model
therefore occurs at {\em late times\/} and on large scales. This unusual
property stands in contrast to braneworld cosmology based on the
Randall--Sundrum model discussed earlier, in which nonstandard behaviour is
encountered at very early times.

\begin{figure}[tbh!]
\centerline{ \psfig{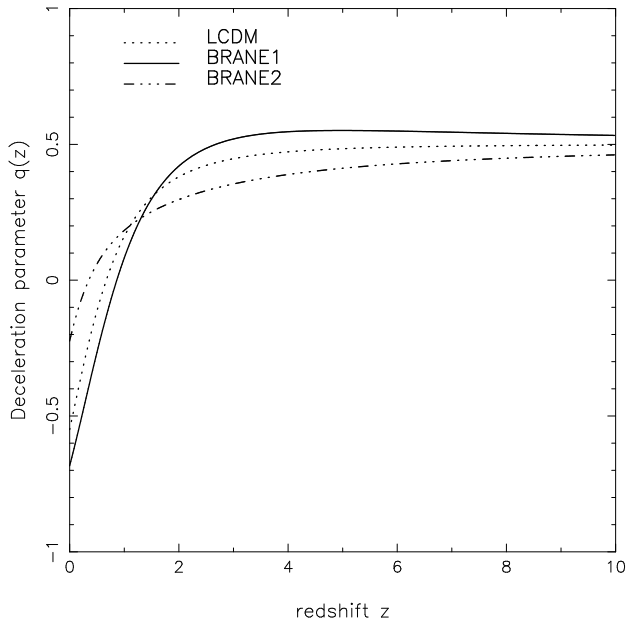} }
\bigskip
\caption{\small The deceleration parameter $q = -{\ddot a}/aH^2$
is shown as a function of redshift for the two braneworld models
and LCDM. In all cases $\Omega_{\rm m} = 0.3$ and the universe is
presently accelerating. Note that the BRANE1 model accelerates at
a {\em faster\/} rate than LCDM which has $\Omega_\Lambda = 0.7$
in the cosmological constant. } \label{fig:acc}
\end{figure}

A spacelike extra dimension leads to dynamical equations on the brane having
the form (\ref{cosmo}) with $\epsilon = 1$. Since observations of the cosmic
microwave background indicate that the spatial curvature is small, we shall set
$\kappa = 0$. We also neglect the radiation density and the `dark radiation'
term $C/a^4$ assuming both to be small today. Equation (\ref{cosmo}) with
$\epsilon = 1$ can be rewritten in terms of the cosmological redshift $z =
a/a_0 - 1$ in the following form:
\begin{equation} \label{hubble1}
{H^2(z) \over H_0^2} = \Omega_{\rm m} (1\!+\!z)^3 +
\Omega_\sigma + \underline{2 \Omega_\ell \pm 2 \sqrt{\Omega_\ell}\,
\sqrt{\Omega_{\rm m} (1\!+\!z )^3 + \Omega_\sigma + \Omega_\ell +
\Omega_{\Lambda_{\rm b}}}} \, ,
\end{equation}
where the dimensionless cosmological parameters
\begin{equation} \label{omegas}
\Omega_{\rm m} =  {\rho_0 \over 3 m^2 H_0^2} \, , \quad \Omega_\sigma = {\sigma
\over 3 m^2 H_0^2} \, , \quad \Omega_\ell = {1 \over \ell^2 H_0^2} \, , \quad
\Omega_{\Lambda_{\rm b}} = - {\Lambda_{\rm b} \over 6 H_0^2}
\end{equation}
correspond to the present epoch. The two signs in (\ref{hubble1}) reflect the
two different ways in which the brane can be a boundary of the bulk; following
\cite{ss02}, we shall refer to the model with `$-$' sign as BRANE1 and the
model with the `$+$' sign as BRANE2. The underlined term highlights the
difference of braneworld dynamics from GR\@.

\begin{figure}[tbh!]
\centerline{ \psfig{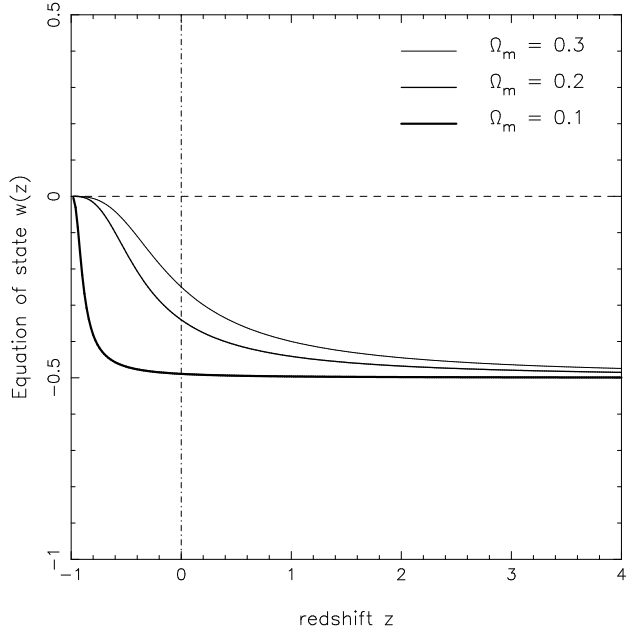} }
\bigskip
\caption{\small The effective equation of state of dark energy in a transiently
accelerating braneworld is shown as a function of redshift for three possible
values of the current matter density. The vertical line corresponds to the
present epoch while the horizontal line refers to $w=0$. The past and future
behaviour of $w(z)$ turns out to be very different: $w(z) \to -1/2$ for $z \gg
1$, while $w(z) \to 0$ for $z \to -1$. Braneworld dark energy disappears in the
future, resulting in the re-emergence of a matter dominated universe. }
\label{fig:state_future}
\end{figure}

Dark energy in the braneworld model has exciting new properties.
This is reflected in Fig.~\ref{fig:acc} and in the expression for
the effective equation of state for dark energy
\cite{ss02}
\begin{equation} \label{state}
w_0 = {2 q_0 - 1 \over 3 \left( 1 - \Omega_{\rm m} \right)} = - 1 \pm
{\Omega_{\rm m} \over 1 - \Omega_{\rm m}} \, {\sqrt{\Omega_\ell \over
\Omega_{\rm m} + \Omega_\sigma + \Omega_\ell + \Omega_{\Lambda_{\rm b}}}} \, .
\end{equation}
Remarkably, the two embeddings of the brane give rise to two distinct
possibilities for dark energy: $w \leq -1$ for the `$-$' sign in (\ref{state}),
and $w \geq -1$ for the `$+$' sign. It should be noted that only the
latter possibility is admitted by conventional models of dark energy including
the cosmological constant and quintessence \cite{ss00}. However, there already
exist indications that $w \leq -1$ may provide a better fit to the existing
supernovae data \cite{caldwell}, a possibility which, if confirmed, could make
a strong case for the braneworld scenario.

Another striking feature of braneworld cosmology is that the current
acceleration epoch (like the radiative and matter-dominated epochs which
preceded it) can be temporary, and could soon come to an end. It is well known
that conventional models of dark energy (cosmological constant, quintessence,
etc.\@) give rise to an eternally accelerating universe with an event horizon,
thus preventing the construction of an $S$-matrix describing particle
interactions within the framework of string or M-theory \cite{horizon}. By
contrast, a negative-tension brane can be `transiently accelerating' as
demonstrated in Fig.~\ref{fig:state_future}. Braneworld models could therefore
succeed where quintessence models have failed, and reconcile an accelerating
universe with the requirements of string/M-theory \cite{Kofinas,ss02}.

\bigskip

{\em Acknowledgments\/}: The authors acknowledge support from the
Indo-Ukrainian program of cooperation in science and technology.  The work of
Yu.~S. was also supported in part by the INTAS grant for project No.~2000-334.

\end{document}